\documentclass[runningheads]{llncs}

\usepackage{booktabs}
\usepackage{multirow}
\usepackage{amsmath, amssymb}
\usepackage{marvosym}  
\usepackage{hyperref}
\hypersetup{
    pdfborder = {0 0 0}
}
\usepackage{graphicx}
\usepackage{amsmath}
\usepackage{amssymb}
\usepackage{booktabs}
\usepackage{multirow}
\usepackage{paralist}
\usepackage{array}
\usepackage[normalem]{ulem}
\usepackage[accsupp]{axessibility}
\usepackage{soul}
\usepackage{url}
\usepackage{color, colortbl}
\usepackage{relsize}
\usepackage{subcaption}
\usepackage{xspace}

\usepackage[capitalize]{cleveref}
\crefname{section}{Sec.}{Secs.}
\Crefname{section}{Section}{Sections}
\Crefname{table}{Table}{Tables}
\crefname{table}{Tab.}{Tabs.}

\usepackage[usenames,dvipsnames]{xcolor}

\newcommand{\blockcomment}[1]{}

\newcommand{\FIXME}[1]{\textcolor{red}{#1}\xspace}

\newcommand{\CITEME}[1]{\FIXME{\cite{CITEME}}\xspace}

\makeatother

\usepackage[T1]{fontenc}
\usepackage{graphicx,verbatim}
\begin{document}
%
\title{
    Anatomy-Grounded Synthetic Coronary Angiography for Geometry-Informed Multi-View Matching
}

\titlerunning{GIMM: Geometry-Informed Multi-view Matching}

\author{
    In Kyu Lee\inst{1,2}\thanks{Equal contribution. $\dagger$ Work performed while affiliated with Medipixel, Inc.}$^{\dagger}$\textsuperscript{(\Letter)} \and
    Sumin Seo\inst{1}$^{\star}$ \and
    Jaesik Min\inst{1}
}
\authorrunning{I.K. Lee et al.}
%
\institute{
    Medipixel, Inc., Seoul, Republic of Korea\\
    \email{{\{sumin.seo,jaesik.min\}}@medipixel.io}\\ \and
    University of California San Diego, La Jolla, USA\\
    \email{kl002@ucsd.edu}
}

\maketitle              
\begin{abstract}
Accurate correspondence matching across multiple angiographic views is the prerequisite for 3D coronary reconstruction and interventional guidance. However, the development of robust deep learning models for this task has been stifled by a fundamental data bottleneck. Obtaining ground truth for matching tasks in angiography pairs is prohibitively expensive and hard to scale. To overcome this barrier, we introduce a physically-grounded data generation framework that synthesizes high-fidelity Digital Reconstructed Radiographs (DRRs) from 3D Coronary CT Angiography (CCTA) volumes. Our framework generates dense, highly accurate 3D-to-2D projection labels by simulating realistic C-arm acquisition geometry on patient anatomy at zero human cost. Leveraging this dense supervision, we propose a Geometry-Informed Matching Module (GIMM) that integrates global feature and anatomical structure into correspondence learning. Unlike real angiography where assessment relies on subjective human annotation, our dataset provides 2D correspondence labels with paired images, allowing human-free evaluation. We comprehensively evaluate our method on the proposed CT-derived DRR dataset and demonstrate improvements over other matching baseline models. As the first public benchmark dataset for coronary correspondence matching, we will release our dataset and code in \url{http://github.com/medipixel/GIMM}.

\keywords{Local feature matching \and Synthetic data generation \and CT \and Coronary angiography.}


\end{abstract}
\section{Introduction}
Multi-view coronary angiography is routinely used for diagnosis and interventional planning in coronary artery disease. Accurate cross-view correspondence is a prerequisite for 3D reconstruction, quantification, and geometry-aware guidance during percutaneous coronary interventions (PCI)~\cite{girasis2013advanced,nishi2017comparison}. While classical geometric reconstruction pipelines rely on handcrafted features and epipolar constraints, 
recent advances in deep learning matching models have demonstrated strong performance in natural image domains~\cite{edstedt2024roma,sarlin2020superglue}. 

However, transferring these advances to medical imaging remains difficult. Unlike natural images, X-ray projections suffer from overlapping anatomical structures and an inherent lack of 3D ground truth. Traditional feature correspondence estimation, which is fundamental to visual localization and pose estimation~\cite{leroy2024grounding,lowe1999object,sarlin2020superglue,schonberger2016structure,weinzaepfel2019visual}, relies heavily on appearance similarity rather than complex projective geometry. For example, SuperGlue~\cite{sarlin2020superglue} introduced a graph neural network-based matching framework that refines sparse keypoint correspondences through attention-based context aggregation.
Subsequently, LoFTR~\cite{sun2021loftr} eliminated the need for explicit keypoint detection by adopting a detector-free, coarse-to-fine Transformer architecture for dense feature matching.
While effective in natural domain, they struggle in coronary angiography due to repetitive tubular structures, severe viewpoint variations, and projection-induced ambiguities, making dense correspondence estimation particularly challenging due to an inherent lack of 3D ground truth.

Establishing geometry-grounded, pixel-level correspondences requires expert annotation, which is expensive and yields only sparse labeling. Moreover, due to the projective nature of X-ray imaging, a 2D point corresponds to a 3D ray, making ground-truth verification inherently ambiguous.
To address this limitation, we propose a physically grounded data generation framework that synthesizes Digital Reconstructed Radiographs (DRRs)~\cite{siddon1985fast,wu2023towards} directly from patient-specific 3D Coronary CT Angiography (CCTA). By simulating clinically realistic C-arm acquisition geometries, our framework produces multi-view angiographic projections that preserve complex coronary anatomy. Crucially, this approach generates data from actual patient anatomy and includes precise 3D-to-2D correspondence labels, entirely eliminating the need for manual annotation.

Leveraging this dense anatomical supervision, we introduce a Geometry-Informed Matching Module (GIMM), which integrates geometric priors to resolve angiographic ambiguities. Our module improves matching in a coarse-to-fine manner: coarsely by incorporating global view information via C-arm angles to orient the network macroscopically, and finely by utilizing epipolar information to constrain and guide localized pixel-level matches. This dual integration of global structure and anatomical consistency significantly improves robustness across challenging view configurations.
Unlike real angiography, where evaluation relies on subjective human annotation~\cite{kim2024effective}, our dataset provides complete 3D ground truth, enabling objective and automated assessment of correspondence quality. We comprehensively evaluate our approach against other deep learning based matching models using both 2D and 3D distance errors. This work establishes the first public benchmark dataset for coronary correspondence matching. 

In summary, this work makes three core contributions:
(1) A physically grounded CT-derived DRR dataset with dense 3D-2D correspondence labels; (2) A geometry-informed matching module that leverages C-arm angle information and epipolar constraints for robust coarse-to-fine correspondence learning; (3) A comprehensive evaluation protocol enabling human-free correspondence assessment. 
\section{Method}
\subsection{CT-Derived DRR Dataset}\label{sec:ct_drr}
\begin{figure}[t!]
    \centering
    \includegraphics[width=1.0\textwidth]{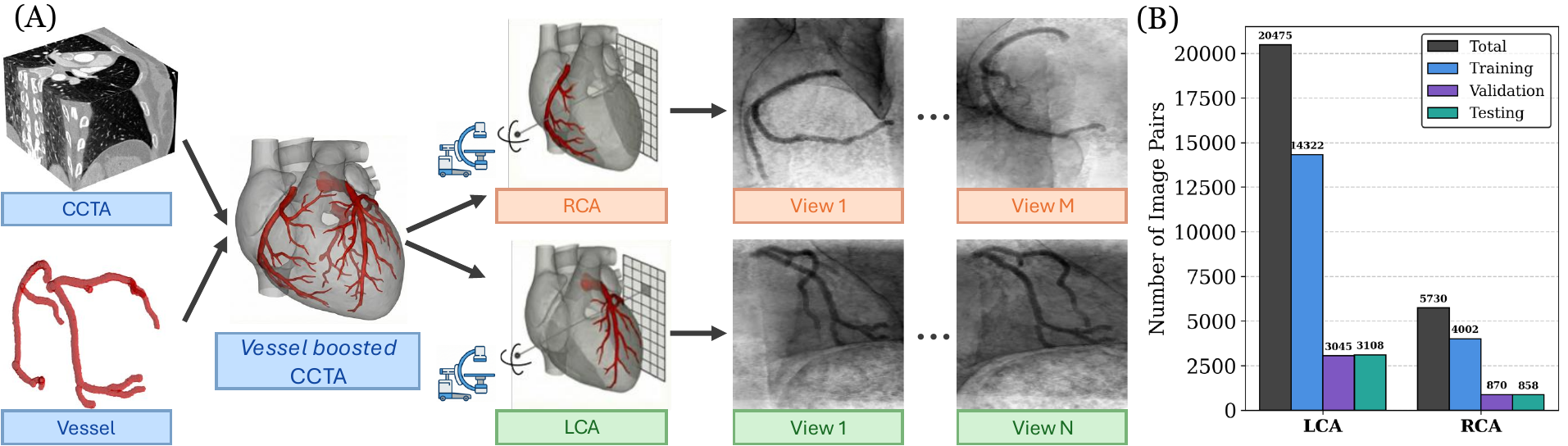}
    \caption{(A) Overview of the CT DRR pipeline. (B) CT DRR data distribution.}
    \label{fig:ct_drr}
\end{figure}

To enable scalable and anatomically consistent supervision for multi-view correspondence learning, we construct a CT DRR dataset from CCTA with vessel segmentation masks, as illustrated in Fig.~\ref{fig:ct_drr} (A).
Given a CCTA volume \(V(x)\), DRRs are generated using a forward projection model that simulates C-arm X-ray acquisition. For each detector pixel, intensity is computed by integrating attenuation coefficients along each X-ray ray according to Beer-Lambert law~\cite{kak2001principles}:
$
I(u)=I_0 \exp(-\int_{R(u)}\mu(x)dx),
$
where \(R(u)\) denotes the ray corresponding to detector coordinate \(u\), and \(\mu(x)\) is the attenuation coefficient at location \(x\).

To approximate contrast enhanced clinical angiography, we modulate the attenuation coefficients using anatomical masks. The coronary tree is separated into isolated right coronary artery (RCA) and left coronary artery (LCA) volumes. By amplifying the attenuation values within these masks and identifying bone structures via intensity thresholding, we realistically simulate contrast agent injection and radiographic anatomy. To reflect standard diagnostic protocols, we simulate clinical C-arm projection angles (LAO/RAO and cranial/caudal). We generate 4 routine views for RCA and 7 for  LCA per patient scan. Multi-view pairs are then constructed from all unordered combinations within each isolated system, yielding 6 RCA pairs and 21 LCA pairs per scan.

\subsection{Geometry-Informed Matching Module}
\begin{figure}[t!]
    \centering
    \includegraphics[width=1.0\textwidth]{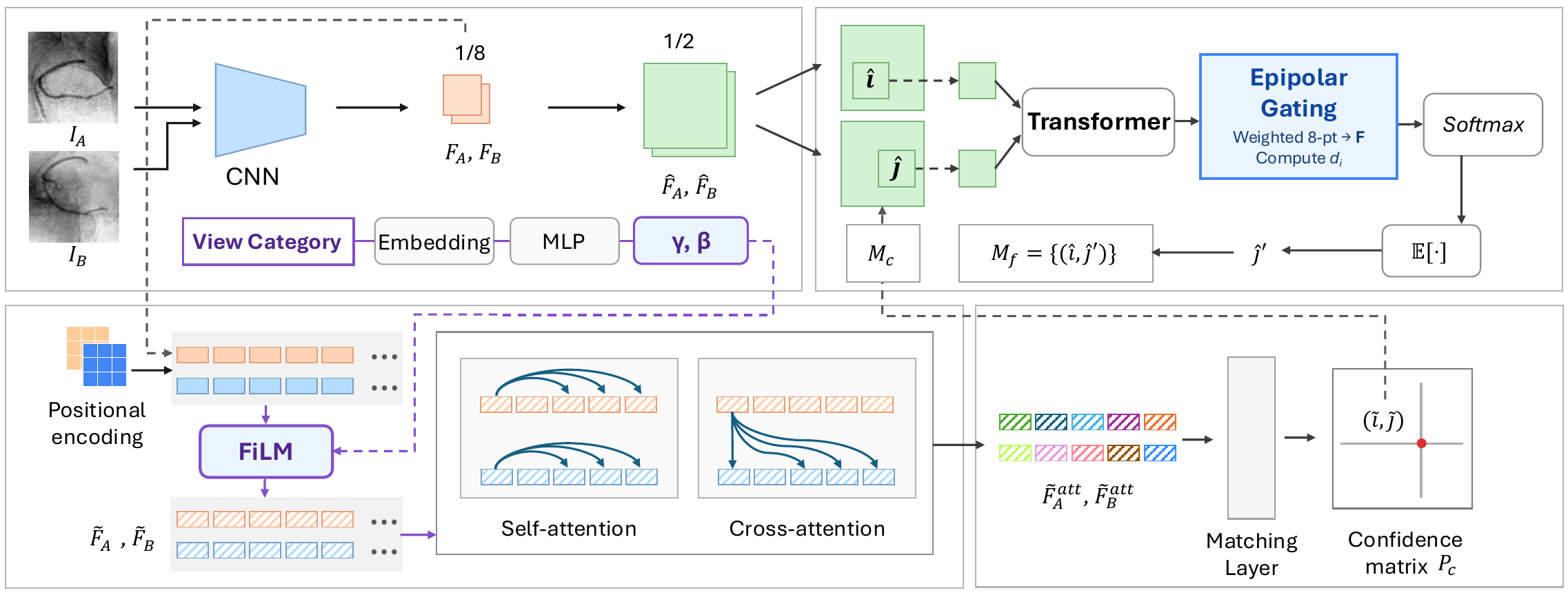}
    \caption{Overall architecture of GIMM.}
    \label{fig:overall_architecture}
\end{figure}

To leverage the dense anatomical supervision of our dataset, we propose a Geometry-Informed Matching Module (GIMM) integrated into a coarse-to-fine correspondence matching framework, as depicted in Fig.~\ref{fig:overall_architecture}.
Throughout the paper, $F_k$, $\hat{F}_k$, and $\tilde{F}_k$ denote the coarse-level, fine-level, and class-modulated features of image $k$, respectively.
We denote the predicted correspondence pair by $(\hat{i},\hat{j}')$, and the coarse and fine correspondence sets by $\mathcal{M}_c$ and $\mathcal{M}_f$.
The coarse stage utilizes a shared ResNet backbone~\cite{he2016deep} and Transformer~\cite{tangquadtree} to establish globally consistent coarse matches. The fine stage then processes local feature patches around each coarse match and estimates the final sub-pixel correspondence as the expected location over the local matching probability distribution, i.e., $\hat{j}' = \mathbb{E}[j']$, yielding the final fine-level matches $\mathcal{M}_f={(\hat{i},\hat{j}')}$.

Our GIMM explicitly enhances this baseline architecture. At the coarse level, a class conditioning mechanism injects acquisition-view priors into the feature representations to guide view-aware learning. At the fine level, an epipolar gating mechanism restricts matching range to geometrically plausible regions. Together, these modifications promote anatomically and geometrically consistent multi-view matching.

\subsubsection{Coarse-to-Fine Matching for Paired Correspondences.}
Following~\cite{sun2021loftr}, we adopt a coarse-to-fine matching strategy.
In standard matching, the coarse correspondence matrix is built by warping grid points with depth and pose and selecting mutual nearest neighbors. 
For coronary angiography, reliable depth and pose are unavailable, so we replace the warp-based supervision with point-pair correspondence labels. 
Specifically, we use the point-pair labels to directly supervise both the coarse-level mapping and the fine-level 2D pixel correspondences.
We applied the same modification to all baselines to ensure a fair comparison. 

\subsubsection{Class-Conditioning Module.}\label{sec:class_cond}
Coronary angiographic projections acquired under different C-arm configurations exhibit distinct global anatomical layouts. For example, LAO and RAO views emphasize different vessel segments and overlap patterns. We hypothesize that explicitly incorporating view class information can provide a useful global prior during coarse correspondence establishment.

Let \(I_k\) denote the input image from view \(k\), and let \(F_k \in \mathbb{R}^{N\times C}\) denote the corresponding coarse-level feature tokens extracted by the backbone, where \(N\) is the number of spatial tokens and \(C\) is the feature dimension. Each image is associated with a discrete view class label \(c_k\in\{1,...,K\},\) corresponding to clinically defined projection categories.

We encode the view class label using a learnable embedding vector, \(e_k \in \mathbb{R}^d \) with \(d\) embedding dimension. 
The embedding is then mapped to feature modulation parameters using two learnable functions, where \(\gamma_k,\beta_k \in \mathbb{R}^C\).
We adopt a Feature-wise Linear Modulation (FiLM)~\cite{perez2018film} to inject the global prior into coarse features:
\begin{equation}
    \tilde{F_k}=\gamma_k \odot F_k + \beta_k
\end{equation}
where \(\odot\) denotes channel-wise multiplication. This modulation adaptively rescales and shifts feature channels based on the projection class, allowing the model to condition its matching behavior on view-specific anatomical configurations. 

The modulated features \(\tilde{F}_k\) are subsequently passed to the coarse Transformer module, producing Transformer-processed features \(\tilde{F}^{\mathrm{att}}_k\). By integrating acquisition-view information directly into feature representations, the proposed class-conditioning module encourages globally coherent matching across anatomically distinct projections.

\subsubsection{Epipolar Geometry Module.}
During the fine matching phase, we apply epipolar gating to the similarity matrix. 
Given tentative correspondences \\$\{x_i, y_i, x_i', y_i'\}^N_{i=1}$ with confidence weights $w_i$, we estimate a fundamental matrix ${\mathbf{F}}\in\mathbb{R}^{3\times3}$ by a weighted 8-point algorithm~\cite{hartley1997defense,yi2018learning}. 
Each correspondence defines a row
$a_i = [x'_ix_i, x'_iy_i, x'_i, y'_ix_i, y'_iy_i, y'_i, x_i, y_i, 1],$
and stacking all rows yields ${\mathbf{A}} \in \mathbb{R}^{N \times 9}$. 
We solve
\begin{equation}
\min_f \| {\mathbf{W}} {\mathbf{A}} f \|^2, 
\quad 
{\mathbf{W}} = \mathrm{diag}(\sqrt{w_i}),
\end{equation}
where $f \in \mathbb{R}^9$ is the column-wise vectorization of the normalized fundamental matrix $\tilde{\mathbf{F}}$.
For numerical stability, we normalize points with weighted centroids
$\mu _0$, $\mu _1$, 
and weighted mean distances
$d_0$, $d_1.$
The normalization transforms are 
\begin{equation}
    \mathbf{T}_0 =
    \begin{bmatrix}
    s_0 & 0 & -s_0 \mu_0^x \\
    0 & s_0 & -s_0 \mu_0^y \\
    0 & 0 & 1
    \end{bmatrix},
    \quad
    \mathbf{T}_1 =
    \begin{bmatrix}
    s_1 & 0 & -s_1 \mu_1^x \\
    0 & s_1 & -s_1 \mu_1^y \\
    0 & 0 & 1
    \end{bmatrix},
\end{equation}
where $
s_0 = {\sqrt{2}}/{d_0}, 
s_1 = {\sqrt{2}}/{d_1}
$. 
We compute the weighted solution on normalized points to obtain $\tilde{\mathbf{F}}$, then enforce $\text{rank}(\tilde{\mathbf{F}}) = 2$ by singular value decomposition, 

and denormalize $
\mathbf{F} = \mathbf{T}_1^\top \tilde{\mathbf{F}} \mathbf{T}_0.
$ 
The estimated ${\mathbf{F}}$ is used to compute the symmetric epipolar distance $d_i$ for each predicted correspondence.
A binary inlier mask is applied based on a distance threshold $\delta$.
The matching confidence is then gated, so that geometrically inconsistent correspondences are explicitly removed rather than softly attenuated.
We also consider two relaxed variants in our ablation studies:
\emph{soft gating} $\exp(-d_i/\tau^2)$ and
\emph{logit gating} $\sigma(-d_i/\tau^2).$

\section{Experiments}
\begin{table}[t]
\centering
\caption{Main result on the synthetic CT-derived DRR dataset (top-$K = 20$).}
\label{tab:synthetic}
    \begin{tabular}{l|ccc|c}
    \toprule
    & & 2D & & 3D \\
    Method & Avg. Dist (px) $\downarrow$ & Prec@3px $\uparrow$ & Prec@5px $\uparrow$ & Avg. Dist (mm) $\downarrow$\\
    \midrule
    SuperGlue~\cite{detone2018superpoint,sarlin2020superglue}  & 7.249 &  0.425 & 0.691 & 4.251\\
    ASpanFormer~\cite{chen2022aspanformer}  & 5.678 & 0.515 & \textbf{0.878} & 1.478 \\
    LoFTR~\cite{sun2021loftr}  & 5.394 & 0.505 & 0.772 & 1.395 \\
    QuadTree~\cite{tangquadtree}  & 5.102 & 0.536 & 0.795 & 1.238 \\
    \midrule
    \textbf{Ours} & \textbf{4.738} & \textbf{0.547} & 0.813 & \textbf{1.171} \\
    \bottomrule
    \end{tabular}
\end{table}
\begin{table}[t]
\centering
\caption{
Ablation study on the proposed components (top-$K = 20$).
}
\label{tab:ablation}
\begin{tabular}{l|cc|c}
\toprule
Method 
& Class Cond.  
& Epi. Gating 
& 2D Dist (px) $\downarrow$ \\
\midrule
Quadtree~\cite{tangquadtree} 
&  &  & 5.102 \\

+ Class Conditioning
& \checkmark &  & 5.038 \\

+ Epipolar Gating (Soft) 
& \checkmark & \checkmark & 5.001\\

+ Epipolar Gating (Logit) 
& \checkmark & \checkmark & 4.887\\

\midrule

\textbf{Ours} 
& \checkmark & \checkmark & \textbf{4.738}\\

\bottomrule
\end{tabular}
\end{table}

\subsection{CT-DRR Dataset}
Our synthetic generation pipeline utilizes CCTA scans and vessel annotations from the publicly available ImageCAS dataset~\cite{zeng2023imagecas}. For each view pair, 3D-to-2D projection correspondences are automatically computed using the known 3D vessel geometry and projection matrices, enabling precise and anatomically consistent supervision without manual annotation. The resulting dataset comprises 26,205 multi-view image pairs. To prevent cross-patient information leakage, the data is strictly divided at the patient level (70\% train, 15\% validation, 15\% test), as detailed in Fig.~\ref{fig:ct_drr} (B).

\subsection{Implementation Details}
The proposed model was trained for 20 epochs using a batch size of 12. Network optimization was performed using the AdamW optimizer with a learning rate of $8e-3$. To supervise the training process, we employed a focal loss for the coarse-level matching and an $L_2$ loss for the fine-level localization. For the FiLM layers, we utilized a feature embedding dimension $d$ of 64. Within the epipolar gating module, the epipolar temperature parameter $\tau$ was set to 2.0, and matches were filtered using an epipolar distance threshold $\delta$ of 2.0 pixels.

\subsection{Results}
\subsubsection{Distance Metrics.}
We compute 2D and 3D correspondence errors using ground truth centerlines to strictly penalize anatomical mismatches. For 2D projection error, predictions within the source vessel mask are mapped to the nearest source centerline to calculate the Euclidean pixel distance to the corresponding target centerline. For 3D distance error, the valid matches in both vessel masks are mapped to their nearest 2D centerlines, and the Euclidean distance between their corresponding 3D spatial coordinates is computed.

\subsubsection{Quantitative Results.}
Table \ref{tab:synthetic} summarizes the performance of our method against other matching baselines. For a fair and rigorous comparison, all metrics are reported using the top-20 most confident predicted matches. The proposed method achieves the lowest average distance error of 4.738 pixels, outperforming the strongest baseline, QuadTree (5.102 pixels). Furthermore, our approach demonstrates superior fine-grained localization, yielding the highest precision at thresholds (0.547 at 3 px and 0.813 at 5 px). By explicitly integrating C-arm geometry and epipolar information, our model achieves a 3D distance error of 1.171 mm. Clinically, an error of this magnitude is highly significant, as it falls well below the typical diameter of major coronary arteries.

\begin{figure}[t!]
    \centering
    \includegraphics[width=1.0\textwidth]{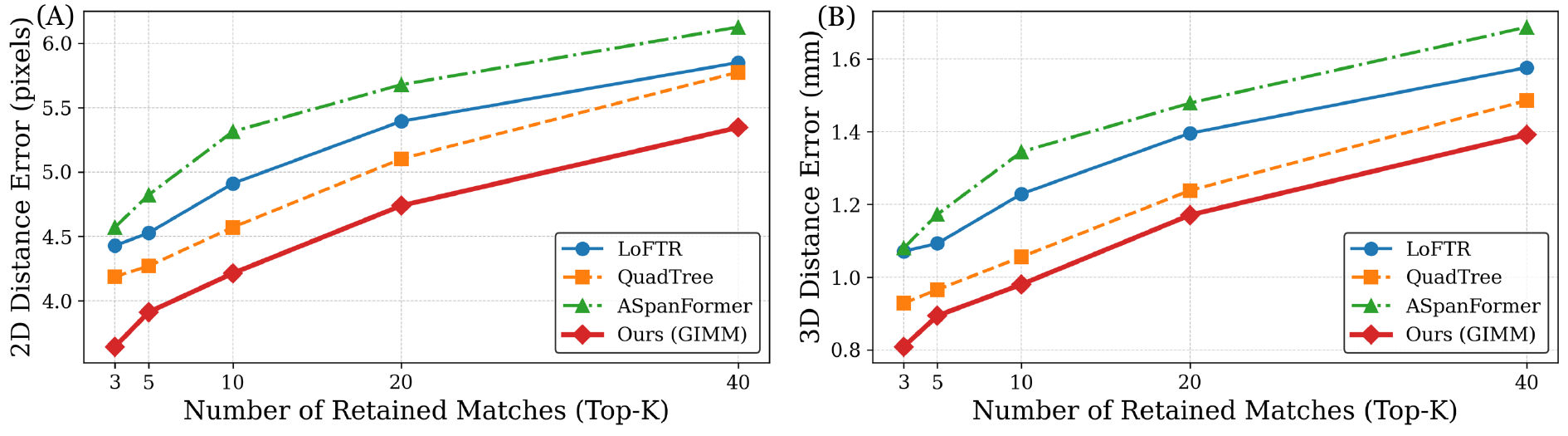}
    \caption{Comparison of our proposed GIMM against baseline methods at various top-$K$: (A) mean 2D error and (B) mean 3D distance.}
    \label{fig:topk}
\end{figure}

\subsubsection{Qualitative Results.}
\begin{figure}[t!]
    \centering
    \includegraphics[width=1.0\textwidth]{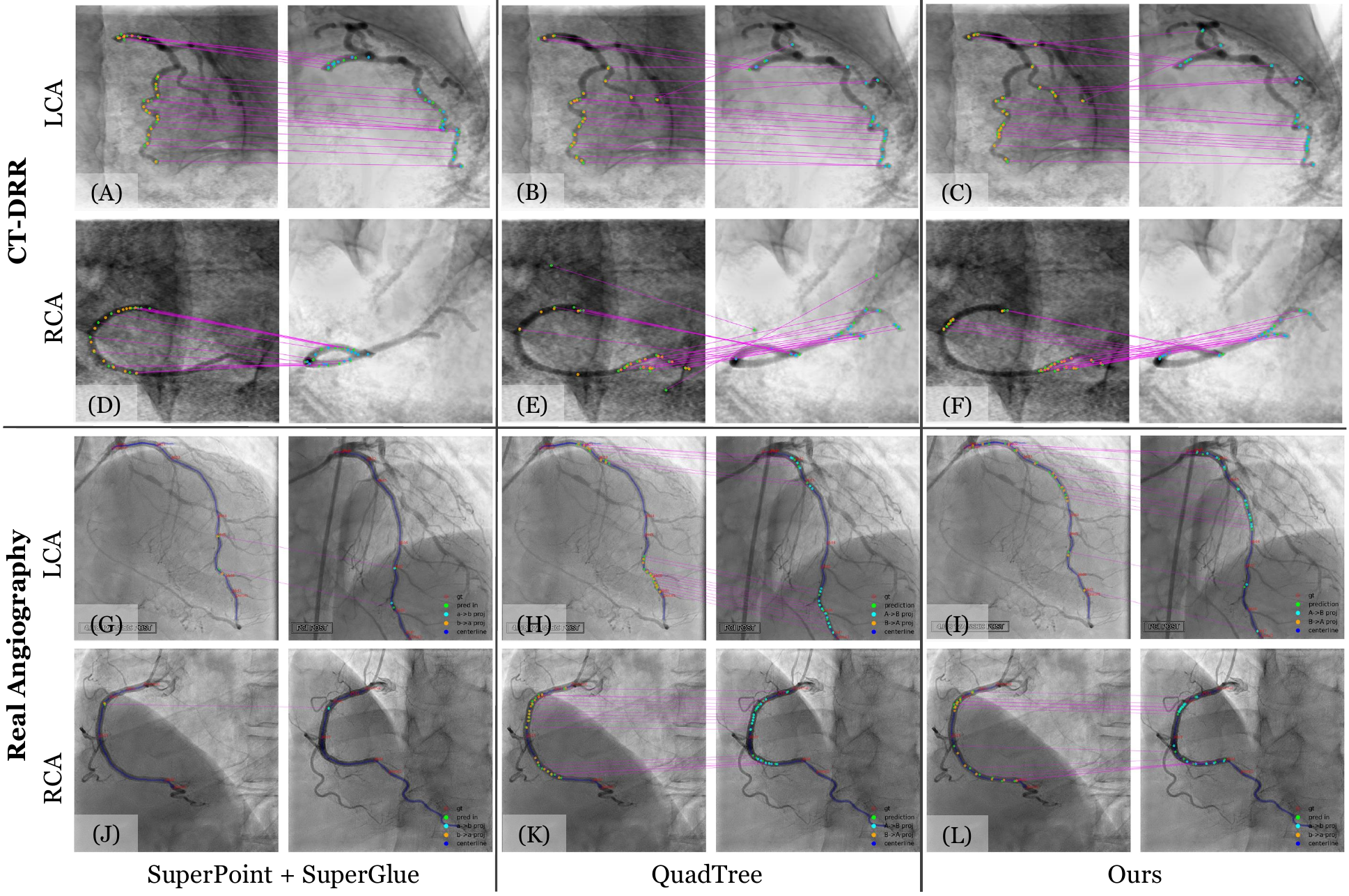}
    \caption{Qualitative results. Predicted matches are shown in green, with magenta lines indicating paired correspondences. Orange and cyan markers represent right-to-left and left-to-right projections, respectively.}
    \label{fig:qualitative}
\end{figure}
Fig.~\ref{fig:qualitative} shows qualitative improvements in correspondence matching over SuperGlue and QuadTree. Beyond higher match density, our method preferentially selects anatomically structured points such as vessel center line-aligned regions and bifurcations, resulting in geometrically coherent matching in both LCA and RCA synthetic images. Despite the domain gap between synthetic and real angiography, our method maintains coherent, distributed structural alignment and demonstrates robustness to appearance variation. 

\subsection{Ablation Studies}
\subsubsection{Robustness Across Confidence Thresholds (Top-$K$ Analysis).} Standard recall metrics are inherently ill-posed for continuous tubular structures due to discrete feature grids. To evaluate the trade-off between match quantity and geometric quality, we compare top-$K$ predictions, as shown in Fig.~\ref{fig:topk}. As $K$ increases, models must output lower-confidence matches. GIMM consistently outperforms all baselines across every threshold. This confirms our module reliably maintains high geometric accuracy across the entire confidence spectrum, rather than artificially inflating precision by outputting only highly certain predictions.

\subsubsection{Effectiveness of Proposed Components.}
To validate the individual contributions of GIMM, we conduct a step-wise ablation study starting from QuadTree (Table \ref{tab:ablation}). Introducing Class Conditioning yields a reduction in 2D distance error by $0.064$. Integrating known C-arm projection geometry via Epipolar Gating further improves performance. While applying a soft epipolar mask lowers the error to $5.001$ px, applying the constraint directly at the logit level sharply improves precision to $4.887$ px. Finally, optimally integrating both semantic conditioning and hard epipolar gating yields the best overall 2D error of $4.738$ px, demonstrating the synergy of these proposed modules.

\section{Conclusion}
We present the first benchmark dataset for dense correspondence matching in coronary angiography. We further propose a view-aware global conditioning module with a geometry-aware module that effectively leverages the underlying 3D structural information for dense matching labels.
Extensive experiments demonstrate that our approach consistently outperforms existing methods, with qualitative results further illustrating high-quality and anatomically consistent correspondences. 
While our qualitative results confirm that models trained on this synthetic data successfully generalize to real clinical angiograms, a rigorous quantitative evaluation on real data requires prohibitive manual annotation efforts and falls outside the scope of this study.

\begin{credits}
\subsubsection{\ackname}
The retrospective collection of clinical angiography data from patients at Soonchunhyang University Cheonan Hospital (Cheonan, Republic of Korea; before 2024) was approved by the Institutional Review Board.

\subsubsection{\discintname}
Sumin Seo and Jaesik Min are current employees of Medipixel, Inc. In Kyu Lee was previously employed by Medipixel, Inc. and is currently affiliated with UC San Diego. 

\end{credits}

%
%
%
\bibliographystyle{splncs04}
\bibliography{main}





\end{document}